\documentclass[journal,twoside]{IEEEtran}
\usepackage{cite}
\usepackage{amsmath,amssymb,amsfonts}
\usepackage{algorithm}
\usepackage{algpseudocode}
\usepackage{graphicx}
\usepackage{textcomp}
\def\BibTeX{{\rm B\kern-.05em{\sc i\kern-.025em b}\kern-.08em
    T\kern-.1667em\lower.7ex\hbox{E}\kern-.125emX}}
\newcommand{\br}{\boldsymbol{r}}
\begin{document}
\title{$\mu$NeuFMT: Optical-Property-Adaptive Fluorescence Molecular Tomography via Implicit Neural~Representation}
\author{Shihan Zhao, Jianru Zhang, Yanan Wu, Linlin Li, Siyuan Shen, Xingjun Zhu,
Guoyan Zheng, \IEEEmembership{Member IEEE}, Jiahua Jiang, and Wuwei Ren
\thanks{
The work was supported by Science and Technology Commission of Shanghai Municipality under Grant 25JS2830300 (Wuwei Ren), National Natural Science Foundation of China under Grants 62105205 (Wuwei Ren), and Shanghai Clinical Research and Trial Center (Wuwei Ren). \emph{(Corresponding author: Wuwei Ren.)}}
\thanks{Shihan Zhao and Jianru Zhang contributed equally to this work.}
\thanks{Shihan Zhao, Yanan Wu, Linlin Li and Siyuan Shen are with the School of Information Science and Technology, ShanghaiTech University, Shanghai 201210, China (e-mail: zhaoshh, wuyn1, lill, shensy2023@shanghaitech.edu.cn).}
\thanks{Jianru Zhang and Jiahua Jiang are with the School of Mathematics, University of Birmingham, Edgbaston, B15 2QN, United Kingdom (e-mail: jxz389@student.bham.ac.uk; j.jiang.3@bham.ac.uk).}
\thanks{Xingjun Zhu is with the School of Physical Science and Technology, ShanghaiTech University, Shanghai 201210, China (e-mail: zhuxj1@shanghaitech.edu.cn).}
\thanks{Guoyan Zheng is with the Institute of Medical Robotics, School of Biomedical Engineering, Shanghai Jiao Tong University, Shanghai 200240, China (e-mail: guoyan.zheng@sjtu.edu.cn)}
\thanks{Wuwei Ren is with the School of Information Science and Technology, ShanghaiTech University, Shanghai 201210, China, and also with State Key Laboratory of Advanced Medical Materials and Devices, ShanghaiTech University, Shanghai 201210, China (e-mail: renww@shanghaitech.edu.cn).}
}

\maketitle

\begin{abstract}
    Fluorescence Molecular Tomography (FMT) is a promising technique for non-invasive 3D visualization of fluorescent probes, but its reconstruction remains challenging due to the inherent ill-posedness and reliance on inaccurate or often-unknown tissue optical properties. While deep learning methods have shown promise, their supervised nature limits generalization beyond training data. To address these problems, we propose $\mu$NeuFMT, a self-supervised FMT reconstruction framework that integrates implicit neural-based scene representation with explicit physical modeling of photon propagation. Its key innovation lies in jointly optimize both the fluorescence distribution and the optical properties ($\mu$) during reconstruction, eliminating the need for precise prior knowledge of tissue optics or pre-conditioned training data. We demonstrate that $\mu$NeuFMT robustly recovers accurate fluorophore distributions and optical coefficients even with severely erroneous initial values (0.5$\times$ to 2$\times$ of ground truth). Extensive numerical, phantom, and in vivo validations show that $\mu$NeuFMT outperforms conventional and supervised deep learning approaches across diverse heterogeneous scenarios. Our work establishes a new paradigm for robust and accurate FMT reconstruction, paving the way for more reliable molecular imaging in complex clinically related scenarios, such as fluorescence guided surgery.
\end{abstract}

\begin{IEEEkeywords}
fluorescence molecular tomography, image reconstruction, optical properties, implicit neural representation, fluorescence guided surgery
\end{IEEEkeywords}

\newpage
\section{INTRODUCTION}
\label{sec:1}

Fluorescence Molecular Tomography (FMT) has emerged as a promising molecular imaging modality, enabling non-invasive 3D visualization of fluorescent probe distribution in highly turbid media \cite{song2025advances}.
Owning to its high sensitivity, cost-effectiveness, and non-ionizing radiation, FMT has found broad applications in preclinical research and clinical practice, e.g., brain imaging \cite{zhu2023near}, \emph{in vitro} cell culture monitoring \cite{ozturk2020high} and tumor detection \cite{cai2020nir}.
FMT reconstructs the internal fluorescence distribution by measuring the photon distribution on the object surface and minimizing the discrepancy between real measurements and the simulated results.
A canonic framework of FMT reconstruction employs an iterative optimization strategy with a forward model — typically based on the Radiation Transfer Equation (RTE) or the diffusion equation (DE) — and an inverse model that refines fluorescence distribution prediction until convergence \cite{arridge2009optical, wang2007biomedical}.
However, FMT reconstruction is inherently ill-posed due to insufficient surface measurements and strong photon scattering effect within tissues.
Development of high-precision FMT reconstruction algorithms is highly attractive for further improvement of the spatial resolution and detection depth of this technology, which ultimately broaden the biological applications in drug discovery \cite{stuker2011fluorescence} and clinical use such as fluorescence guided surgery \cite{mieog2022fundamentals, wang2023fluorescence}.

To address the ill-posed nature of FMT reconstruction, regularization techniques that incorporate prior knowledge are applied. Conventional approaches include Tikhonov ($L_2$~norm) regularization for smoothing the results \cite{zhu2014comparison, yi2013reconstruction}, and sparse-promoting regularization ($L_1$, $L_p$ norm) or total variation~(TV) regularization for enhancing the spatial resolution \cite{shi2013efficient, zhu2014nonconvex}.
Recent advancements in regularization techniques, such as the solution-decomposition algorithm, utilize an advanced stochastic prior model that improves the differentiation of fluorescence targets from background interference, thereby enhancing reconstruction quality\cite{zhang2025high}.
While these methods stabilize solutions, their reliance on time-consuming iterative optimization and idealized assumptions about optical parameters often results in slow convergence and limited accuracy in heterogeneous tissues.
Deep learning (DL) methods, particularly en-decoder-based neural networks \cite{huang2019fast, guo20193d, zhang2021uhr, cheng2022encoder, zhang2022multi, zhang2023robust, zhang2025pah2t} and graph-based networks \cite{meng2020k, li2020reconstruction, wang2022graph}, have recently been used in FMT reconstruction by learning direct mappings from boundary measurements to fluorescence distributions.
These approaches encode physical priors within network parameters, enabling overwhelmingly rapid inference. However, their dependence on supervised training with large, scenario-specific datasets restricts generalizability across variable experimental conditions (e.g., tissue object geometry and optical properties), leading to significant accuracy degradation when the object properties deviate from training assumptions.

Recently, implicit neural representation (INR) models have emerged as a powerful alternative for medical image reconstruction.
By mapping arbitrary spatial coordinates to certain unknown physical quantities via a coordinate-based multilayer perceptron network (MLP), INRs achieve continuous 3D representations, self-supervised training, and memory efficiency.
Originally popularized in computer vision and medical imaging, INRs have demonstrated remarkable success in modalities such as computed tomography (CT), magnetic resonance imaging (MRI) \cite{sun2021coil, lozenski2022memory, reed2021dynamic, shen2022nerp, liu2022recovery}, and notably in diffuse optical tomography (DOT), where INR has been proven to be able to capture a high-resolution, continuous optical absorption map that account for light scattering in complex tissues \cite{shen2024high}.
INR captures continuous spatial information and integrates physical priors makes it  particularly suited for various imaging problems.
In the context of FMT, preliminary work has explored position encoding strategies to enhance spatial resolution and improve the quality of FMT reconstructions \cite{han2025tspe}.
Nevertheless, systematic studies integrating optical diffusion physics within an INR framework remain scarce.
This gap presents an opportunity to harness the strengths of implicit modeling for robust and data-efficient FMT reconstruction.

To bridge this gap, we propose neural-field-based FMT, or NeuFMT, a physics-driven reconstruction framework unifying INR with explicit physical models.
Unlike conventional DL-based FMT reconstruction algorithms that couple explicit fluorescence predictions to implicit physical model, NeuFMT integrates the DE-based forward model in the network loss, enabling self-supervised training without paired data.
The differentiability of the forward model is achieved via finite element method (FEM).
This inversion-free approach ensures rapid, continuous reconstruction across diverse geometries while inherently addressing scattering effects.
Another critical limitation of both traditional and DL methods lies in the presumed optical properties of the object during reconstruction, which largely aggravate the ill-posedness of the inverse problem given  the optical parameters were initially biased.
To overcome this, we further introduce optical-property-adaptive NeuFMT, or $\mu$NeuFMT, an adaptive extension of NeuFMT that dynamically optimizes scattering / absorption coefficients and fluorescence distributions jointly.
By integrating an adaptive module of adjusting optical property during the iterations, $\mu$NeuFMT achieves robust accuracy even when provided with erroneous initial guess of the optical coefficient values, addressing the problem of unrealistic assumption prior to FMT reconstruction.
Both proposed INR-based FMT reconstruction algorithms were thoroughly compared and validated by simulations and phantom experiments along with other methods.
Furthermore, \emph{in vivo} lymph node imaging in mice was performed, further demonstrating the advantages of the proposed methods. 

\section{Methods}
\label{sec:methods}

\subsection{Forward Modeling of FMT}
\label{sec:forward_model}
Model-based FMT reconstruction comprises two fundamental steps: forward modeling and inversion.
The former predicts surface photon density measurements based on an assumed fluorescence distribution.
At the macroscopic scale, we employ the first-order approximation of the radiative transport equation (RTE), known as the diffusion equation (DE) \cite{arridge1999optical}.
The excitation and fluorescence emission processes in continuous-wave (CW) mode form a coupled pair of DEs expressed as \cite{arridge1995photon1, arridge1995photon2}: 
\begin{alignat}{3}
   \label{eq:decex}
   -\nabla \cdot \kappa\, \nabla \phi_x(\br) + \mu_a\, \phi_x(\br) &= q_x(\br),  &\quad \br \in \Omega, \\[4pt]
   \label{eq:decem}
   -\nabla \cdot \kappa\, \nabla \phi_m(\br) + \mu_a\, \phi_m(\br) &= \eta\, C(\br)\, \phi_x(\br), &\quad \br \in \Omega,
\end{alignat}
where $\phi_x$ and $\phi_m$ denote the excitation and emission photon density, $q_x$ is the excitation source, $C(\br)$ is the fluorophore concentration to be reconstructed in the inversion, and $\eta$ collects quantum-yield factors in the CW source term. 
$\kappa$ represents the diffusion coefficient, calculated by \cite{arridge1995photon1}
\begin{equation}
   \label{eq:diffusion}
   \kappa = \frac{1}{3\,(\mu_a + \mu_s')},
\end{equation}
with the absorption coefficient $\mu_a$ and the reduced scattering coefficient $\mu_s'$. In a non-contact setting of FMT, to calculate the value of $\phi_\ell(\br)$ on the object surface $\partial \Omega$ as the measurement, a Robin-type boundary condition is added alongside the equation pair (\ref{eq:decex}\,--\,\ref{eq:decem}):
\begin{equation}
   \label{eq:decbound}
   \phi_\ell(\br) + 2\, \zeta(c)\, \kappa\, \frac{\partial \phi_\ell(\br)}{\partial \nu} = 0, \quad \ell\in\{x,m\},\ \br \in \partial \Omega, 
\end{equation}
where $\zeta(c)$ is the refractive index mismatch parameter that accounts for the light reﬂection on the boundary surface, with $c$ representing the light speed inside the medium. $\nu$ denotes the outward normal on the boundary. 
Notably, to simplified the problem setting, we only consider a homogeneous medium case and the difference of $\mu_a,\,\mu_s'$ values under different wavelengths are neglected. 

\begin{figure} [!ht]
\label{fig:1} 
   \begin{center}
   \begin{tabular}{c}
   \includegraphics[width=\columnwidth]{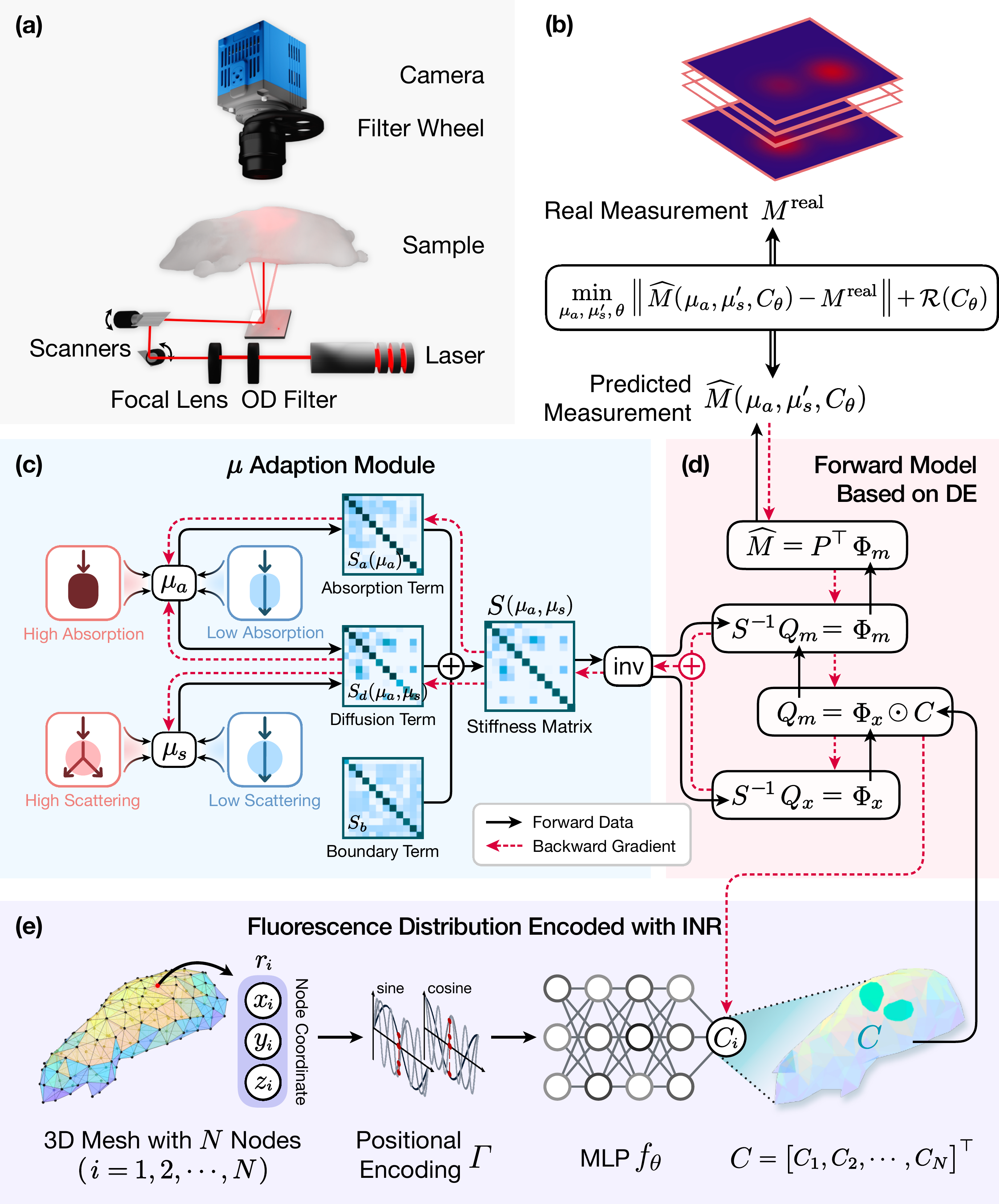}
   \end{tabular}
   \end{center}
   \caption{ 
Conceptual illustration of $\mu$NeuFMT.
\textbf{(a)} The FMT setup: a scanned laser illuminates the sample while a CMOS camera with filter wheel on the opposite side records boundary images. The raster scanning method yields the raw 2D image stack $M^{real}$, serving as the input to the subsequent FMT reconstruction.
\textbf{(b)} The difference between $M^{real}$ and the simulated measurement $\widehat{M}(\mu_a,\mu_s^\prime,C_\theta)$ is iteratively minimized until an optimized value of unknown fluorescence distribution $C$ is found. $R(C)$ is a regularization term and $\mu\in\{\mu_a,\mu_s'\}$ is the intrinsic optical properties consisting of absorption coefficient $\mu_a$ and scattering coefficient $\mu_s'$. 
\textbf{(c)} A $\mu$-adaption module is integrated into the FMT reconstruction. The stiffness matrix $S$ is determined by both $\mu_a$ and $\mu_s'$. 
\textbf{(d)} The FEM-based simulator predicts virtual measurement $\widehat{M}(\mu_a,\mu_s^\prime,C_\theta)$ based on the operator $S^{-1}(\mu)$  and predicted fluorescence distribution $C$. 
\textbf{(e)} Fluorescence distribution encoded in an INR. 3D node coordinate serves as input, while a continuous $C$ map is generated through positional encoding and an MLP network.}
\end{figure} 

\subsection{Finite Element Discretization and System Matrix Construction}
\label{sec:FEM_discrite}
We apply the finite element method (FEM)~\cite{zienkiewicz1977finite} to the diffusion model \eqref{eq:decex}\,-–\,\eqref{eq:decbound}.
Discretizing $\Omega$ with a tetrahedral mesh of $N$ nodes and nodal basis $\{\psi_i\}_{i=1}^N$, we approximate
$\phi_\ell(\br)\approx\sum_{i}\Phi_{\ell,i}\,\psi_i(\br)$ and $q_x(\br)\approx\sum_i(Q_x)_i\,\psi_i(\br)$,
which transforms the continuous-form equations into a linear system for excitation and emission photon propagation:
\begin{align}
   \label{eq:dex}
   S\, \Phi_x &= Q_x \\
   \label{eq:dem}
   S\, \Phi_m &= Q_m
\end{align}
where $S\in \mathbb{R}^{N\times N}$ is the system matrix and $\Phi_\ell,\,Q_\ell\in\mathbb{R}^N$\linebreak are discretized form (coefficient vectors) for $\phi_\ell$ and $q_\ell$, respectively. 
In particular, the excitation field $\phi_x$ interacts with fluorophores $C\in\mathbb{R}^N$ to produce the emission source term
\begin{equation}
   \label{eq:dex2em}
   Q_m = C \odot \Phi_x
\end{equation}
with $\odot$ denoting element-wise multiplication. The equation \eqref{eq:dex2em} physically describes how fluorophores absorb excitation photons and re-emit light at longer wavelengths.
The emission source $Q_m$ then propagates through the same medium to produce the emission photon density $\Phi_m$, which is finally detected at the boundary, which is computed as
\begin{equation}
   \label{eq:dmeas}
   M = P^\top \Phi_m
\end{equation}
where $P$ represents the transfer function, which maps the boundary photon density $\Phi_m$ the measurement data $M$ acquired by the camera. The above linearized forward model provides the physical foundation for the inversion procedure.

Crucially, the SPD FEM stiffness matrix $S(\mu_a,\mu_s')\in\mathbb{R}^{N\times N}$ in \eqref{eq:dex}\,--\,\eqref{eq:dem} encode the optical properties and boundary geometry, generating from a linear combination of matrices corresponding to distinct physical components \cite{ntziachristos2002fluorescence, arridge1995photon1}
\begin{equation}
\label{eq:Sdecomp}
   S(\mu_a, \mu_s^\prime) = c * \left( \mu_a * S_a + \dfrac{1}{3(\mu_a+\mu_s^\prime)} * S_d + S_b \right).
\end{equation}
with $S_a$, $S_d$, and $S_b$ the assembled absorption, diffusion, and Robin-boundary matrices, respectively. This decomposition explicitly reveals the functional dependence of $S$ on the optical parameters $\mu_a$ and $\mu_s^\prime$, which forms the mathematical foundation for our adaptive optimization approach introduced in Subsection \ref{sec:our_method}.

\subsection{INR-based FMT reconstruction: $\mu$NeuFMT / NeuFMT}
\label{sec:our_method}
Unlike traditional voxel-based or mesh-dependent representations, INR provides a continuous, memory-efficient framework for modeling complex physical fields. The basic variant, NeuFMT, learns only the fluorophore $C_\theta$ with fixed optical properties; the proposed adaptive $\mu$NeuFMT jointly optimizes the optical coefficients $(\mu_a,\mu_s')$ with $C_\theta$ for enhancing the reconstruction accuracy and robustness. 
In this section, we describe the $\mu$NeuFMT representation, formulate the learning objective, introduce optical-property adaptation, and summarize the network architecture and implementation.

\subsubsection{INR representation of the unknown fluorescence}
Our $\mu$NeuFMT model the 3D fluorescence field as an INR, i.e., a coordinate-conditioned MLP:
\begin{equation}
\label{eq:C_fomula}  C_\theta(\br) = f_\theta \big(\varGamma(\br)\big),
\end{equation}
where $\br=(x,y,z)\in\Omega\subset\mathbb{R}^3$, $\varGamma:\Omega\to\mathbb{R}^{d}$ is a
high-dimensional positional encoding of spatial coordinates, and
$f_\theta:\mathbb{R}^{d}\to\mathbb{R}$ is an MLP with parameters $\theta$.
This INR formulation \eqref{eq:C_fomula} enables resolution-independent reconstruction and naturally accommodates arbitrary spatial queries without being constrained by the conventional discrete grid structure.

To enhance the network’s ability to represent high-frequency details, which is crucial for resolving complex \emph{in vivo} fluorescence patterns, we apply a sinusoidal positional encoding to each coordinate
component. Let $\gamma:\mathbb{R}\to\mathbb{R}^{2L}$ be
\begin{align}
\label{eq:gamma_encoding}
  \gamma(p) =
  \big(&\sin(2^{0}\pi p),\,\cos(2^{0}\pi p),\,\cdots,\\ \nonumber
         &\sin(2^{L-1}\pi p),\,\cos(2^{L-1}\pi p) \,\big),
  \quad p\in\{x,y,z\},
\end{align}
where $L$ is the number of frequency bands. The 3D position is then encoded as
$\varGamma(\br) = \big(\gamma(x),\,\gamma(y),\,\gamma(z)\big) \in \mathbb{R}^{6L}$,
so that $d=6L$. Feeding $\varGamma(\br)$ into $f_\theta$ enables compact representation of both
low- and high-frequency variations in $C_\theta$.

\subsubsection{Learning objective}
We estimate the fluorescence field $C_\theta$ and, in $\mu$NeuFMT, the optical properties $\{\mu_a,\mu_s'\}$ by minimizing a data-fidelity term plus regularization:
\begin{equation}
   \label{eq:loss}
   \mathcal{L}(\mu_a,\mu_s^\prime,\theta)
   =
   \big\|\widehat{M}(\mu_a,\mu_s',C_\theta) - M^{\mathrm{real}}\big\|_2^2
   + \lambda\,R(C_\theta),
\end{equation}
where $\widehat{M}$ denotes the boundary measurements, and $R(C_\theta)$ is a regularization term applied to the fluorescence distribution.
The hyperparameter $\lambda$ controls the regularization strength, balancing data fidelity with sparsity constraints.
The training of $\mu$NeuFMT / NeuFMT follows a fully self-supervised paradigm that explicitly integrates the DE-based physical model into the optimization loop.  Note that NeuFMT uses fixed optical parameters $(\mu_a,\mu_s^\prime)$, whereas $\mu$NeuFMT updates $\mu$ jointly.

\subsubsection{Alternating optical-property adaptation}
A fundamental innovation of $\mu$NeuFMT lies in formulating the entire photon propagation process—from excitation to emission—as an end-to-end differentiable map. The forward model follows a precise physical sequence from excitation source to surface measurements, as described in Subsection \ref{sec:FEM_discrite}:
\begin{equation}\label{eq:forward-compact}
\widehat{M}(\mu_a,\mu_s^\prime,C_\theta)
=
P^\top S(\mu_a,\mu_s^\prime)^{-1} \big( C_\theta \odot S(\mu_a, \mu_s^\prime)^{-1} Q_x \big),
\end{equation}

We optimize the introduced objective $\mathcal{L}(\mu_a,\mu_s^\prime, \theta)$ in \eqref{eq:loss}, jointly updating network
parameters $\theta$ and optical coefficients $(\mu_a,\mu_s^\prime)$.
The chain in \eqref{eq:forward-compact} contains only differentiable operations, including linear solves,
elementwise products, and linear projections. Thus $\widehat{M}$ is differentiable w.r.t.\ both
$C_\theta$ (hence $\theta$) and $(\mu_a,\mu_s^\prime)$, enabling standard gradient-based training. 
For completeness, we state the chain-rule decomposition of the loss gradient with respect to $(\mu_a,\mu_s^\prime)$. Based on \eqref{eq:forward-compact}, the gradient
splits into an \emph{emission} term and an \emph{excitation} term:
\begin{align}\label{eq:grad-mu-paths}
\frac{\partial \mathcal{L}}{\partial \mu}
=&\underbrace{\frac{\partial \mathcal{L}}{\partial \widehat{M}}\,
\frac{\partial \widehat{M}}{\partial \Phi_m}
\Big(-S^{-1}\,\frac{\partial S}{\partial \mu}\,\Phi_m\Big)}_{\text{Emission pathway}}
\;+\;\nonumber
\\
&\underbrace{\frac{\partial \mathcal{L}}{\partial \widehat{M}}\,
\frac{\partial \widehat{M}}{\partial \Phi_m}\,
\frac{\partial \Phi_m}{\partial Q_m}\,
\frac{\partial Q_m}{\partial \Phi_x}
\Big(-S^{-1}\,\frac{\partial S}{\partial \mu}\,\Phi_x\Big)}_{\text{Excitation pathway}}
\\ &\mu \in \{\mu_a,\mu_s^\prime\} \nonumber.
\end{align}
The only model-specific ingredient of $\mu_a$ and $\mu_s^\prime$ is the Jacobian of $S(\mu_a,\mu_s^\prime)$. According to  \eqref{eq:Sdecomp}, the partial derivatives are  
\begin{align}
\label{eq:dS-dmu-compact}
\frac{\partial S}{\partial \mu_a}
&=c \left[S_a - \frac{1}{3(\mu_a+\mu_s')^2}\,S_d\right], \nonumber\\[4pt]
\frac{\partial S}{\partial \mu_s'}
&=c \left[- \frac{1}{3(\mu_a+\mu_s')^2}\,S_d\right].
\end{align}
This allows $\mu_a$ and $\mu_s^\prime$ to be optimized directly using gradient descent, and then be updated together with $\theta$, bridging the learning process with the underlying physical model.

To mitigate potential coupling effects between $\mu_a$ and $\mu_s'$ during optimization, we propose an alternating update strategy, which periodically updates one optical parameter with the other fixed. The alternating framework largely promotes training stability to avoid oscillations with sufficient computational efficiency. The complete $\mu$NeuFMT algorithm is described in Algorithm \ref{alg:neufmt-ada}. Overall, the alternating $\mu$-adaption module ensures that each optical parameter receives dedicated optimization attention while maintaining the physical consistency of the forward model, yielding more accurate FMT reconstruction results even in the presence of mismatch between the initial guess and ground truth of the optical properties. 

\subsubsection{Network architecture and implementation} 
We adopt MLP for the NeuFMT / $\mu$NeuFMT, comprising 8 fully-connected layers with 512 hidden units each, followed by a final layer with 128 units. ReLU activation functions are employed throughout the network. 
A critical skip connection bridges the input $\varGamma(\br)$ to the 4th hidden layer, facilitating gradient flow and enhancing the ability to learn fine-grained details. 
This architectural choice is particularly suited for FMT applications where accurately reconstructing high-contrast targets against complex backgrounds is essential.

We implement the NeuFMT / $\mu$NeuFMT framework with PyTorch, leveraging its automatic differentiation capability to compute the gradient through both MLP and the physical model. Adam optimizer is used with the initial learning rates of $10^{-4}$ for $\theta$ update, $10^{-3}$ for scattering coefficient and $10^{-5}$ for absorption coefficient. To improve the computational efficiency, a learning rate decay is applied for MLP update. These rates are determined empirically to ensure stable convergence. The $L_1$ regularization parameter $\lambda$ is set to $10^{-6}$ based on cross validation. The training typically requires 2000-20000 iterations to achieve satisfactory reconstruction quality, with the entire process completing within 3\,--\,30 minutes on a single NVIDIA RTX 3090 GPU for problems of varying scales. The FEM discretization is built on TOAST++ toolkit \cite{schweiger2014toast++}, and the FMT simulation is based on a modular software platform STIFT \cite{ren2019smart} in MATLAB (R2024b, MathWorks, MA, US) using a desktop computer (Intel Core i5-12600 CPU @ 3.30 GHz, 128 GB RAM).

\begin{algorithm}[!ht]
\caption{$\mu$NeuFMT}
\label{alg:neufmt-ada}
\begin{algorithmic}[1]
\Require Real measurements $M^{\text{real}}$, light speed $c$, stepsizes $\alpha_\theta,\alpha_a,\alpha_s$, alternation period $T$, total iterations $K$
\State Initialize $\theta,\mu_a,\mu_s^\prime$ ;  
\State Precompute $S_a,S_d,S_b, Q_x, P$ 
\State Compute $S$ via \eqref{eq:Sdecomp}; 
\For{$i=1$ \textbf{to} $K$}
  \State $C\;=\;f_\theta \big(\varGamma(\br)\big)$
  \State Solve $\Phi_x \in \mathbb{R}^{N}$ satisfying $S\,\Phi_x=Q_x$
  \State $Q_m = C \odot \Phi_x$
  \State Solve $\Phi_m \in \mathbb{R}^{N}$ satisfying $S\,\Phi_m=Q_m$
  \State Compute predicted measurements: $\widehat{M} = P^\top \Phi_m$
  \State Compute loss: $\mathcal{L} = \|\widehat{M}-M^{\text{real}}\|_2^2 + R(C)$
  \State Update MLP parameters: $\theta = \theta - \alpha_\theta\,\frac{\partial\mathcal{L}}{\partial \theta}$
  \If{$i \bmod T = 0$}
    \If{$\big\lfloor i/T \big\rfloor$ is even}
      \State Update absorption: $\mu_a = \mu_a - \alpha_a\,\frac{\partial\mathcal{L}}{\partial \mu_a}$
    \Else
      \State Update diffusion: $\mu_s^\prime = \mu_s^\prime - \alpha_s\,\frac{\partial\mathcal{L}}{\partial \mu_s^\prime}$
    \EndIf 
    \State Update $S = c\!\left[\mu_a S_a + \dfrac{1}{3(\mu_a+\mu_s^\prime)} S_d + S_b\right]$; 
  \EndIf
\EndFor
\State \textbf{Output:} fluorescence $C$, optical parameters $\mu_a,\mu_s$
\end{algorithmic}
\end{algorithm}

\subsection{FMT System Setup}
\label{sec:FMT_system}
We devised a continuous-wave (CW) FMT system (SHMOT) \cite{gao2023multifunctional, hu2024simultaneous} as depicted in Fig.\,1\,(a), which incorporates a 16-bit CMOS camera (Edge 4.2, PCO, Germany) with 2048 $\times$ 2048 pixel resolution and a filter wheel (FW102C, Thorlabs, NJ, US) for selective wavelength detection. A tunable laser source (SuperK EXTREME, NKT, Denmark), spanning 400-890 nm, provides fluorescence excitation with a power density of approximately 2 $\text{mW}/\text{mm}^2$ at the sample plane. 
The imaging object can be positioned on a height-adjustable platform, while a galvanometer-driven mirror system (basiCube 10, SCANcube, Germany) mounted aside steers the laser beam to raster-scan either the top or bottom object surface, enabling switchable reflection-/transmission-mode detection. During experiments, the camera exposure time was fixed at 500 ms, with pixel binning applied to improve the signal-to-noise ratio. The acquired raw 2D image data serve as supervision target for NeuFMT / $\mu$NeuFMT reconstruction (Fig.\,1\,(b)). It worth mentioning that our FMT system also integrates a programmable line-scanning illumination, which enables high-precision extraction of the surface geometry with approximated maximum error of 0.1 mm. The extracted surface information is then leveraged as prior knowledge for FEM meshing and improving reconstruction accuracy.

\subsection{Phantom preparation and optical calibration}
In our validation, the slab phantoms
were cast from silicone (SYLGARD\textsuperscript{TM} 184; DOW, CA, USA) to emulate soft-tissue matrix. Optical properties were tuned by dispersing titanium dioxide (TiO$_2$; Colins, Shanghai, China) and carbon black powder (Colins, Shanghai, China) as scattering and absorbing additives, respectively. In all phantoms, TiO$_2$ and carbon black were uniformly distributed. All the designs shown in Fig.\,4\,(a) used the same TiO$_2/$carbon-black mixing ratio, yielding isotropic media with identical absorption and scattering parameters.

\subsection{In Vivo Lymph Node FMT Imaging}
\label{sec:invivo_setup}
Following numerical simulations and phantom experiments, we evaluated the proposed method \emph{in vivo} in a murine lymph node imaging task. 
In cancer diagnostics, lymph node metastasis is a key prognostic indicator of tumor progression. Fluorescence imaging enables rapid screening of sentinel lymph nodes \cite{ji2022biocompatible}.

In our experiment, all animal procedures were approved by the Institutional Animal Care and Use Committee of ShanghaiTech University (Approval No.\ 20211115001). 
5 female BALB/c mice (8~weeks old) were obtained from Shanghai Jihui Laboratory Animal Breeding Co.\ Ltd.\ and housed under a circadian rhythm of 12 hours of day and night alternation. 
Before taking imaging experiments, fur was removed over the whole body. 
For FMT reconstruction, we assumed an optically homogeneous medium with properties calibrated during the excitation step with $(\mu_a = \text{0.0055 mm}^{-1},\ \mu_s = \text{1.1 mm}^{-1})$. 

We followed a three-stage protocol—fluorescent-dye injection, imaging, and \emph{ex vivo} validation.
Prior to FMT imaging, Cy5-saline solution $(\text{1.2146}\,\mu\text{mol\,mL}^{-1},\,40\,\mu\text{L})$ was injected subcutaneously into the left forepaw to drive lymphatic drainage toward the sentinel lymph nodes and create a high-intensity target.
Next, follow-up FMT measurements were acquired using the system setup in Subsection \ref{sec:FMT_system}.
Finally, after data collection, ex vivo verification was performed using a fluorescent stereoscopic microscope (MVX10, Olympus, Japan; 6.3$\times$ objective).
Illumination was provided by a height-adjustable 630 nm LED at $\text{20 mW\,cm}^{-2}$.
White-light and Cy5-channel fluorescence images were recorded using filters matched to the Cy5 spectrum, and the lymph nodes were identified, excised, and imaged to validate the \emph{in vivo} findings.

\begin{figure*} [htbp]
\label{fig:2} 
   \begin{center}
   \begin{tabular}{c}
   \includegraphics[width=12.2cm]{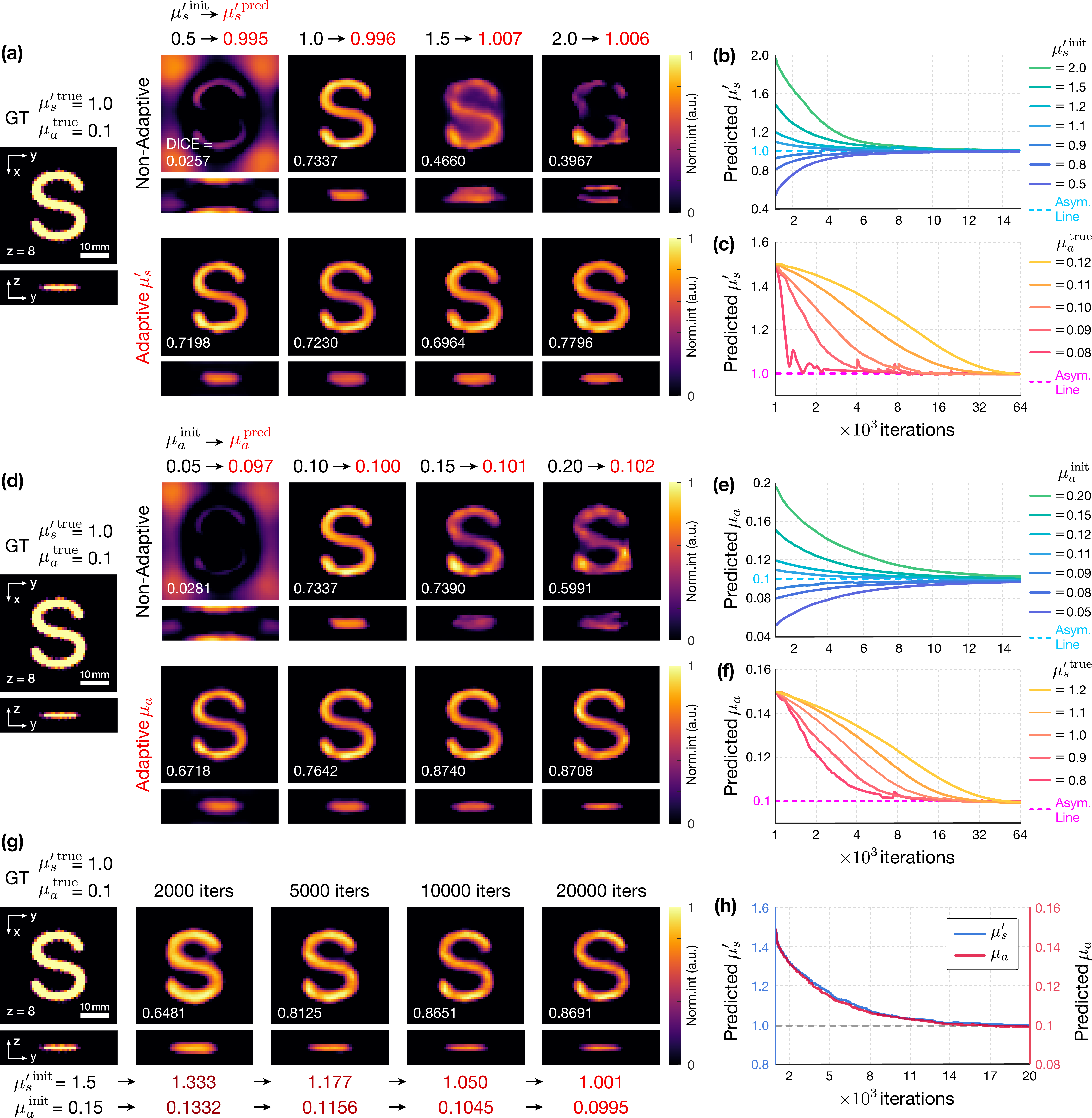}
   \end{tabular}
   \end{center}
   \caption{ 
Ablation test for the optical-property-adaption ($\mu$-adaption) module in $\mu$NeuFMT.
\textbf{(a-c)} Effective test for the correction of $\mu_s'$.
\textbf{(a)} For a homogeneous phantom simulation with a ‘S’-shaped fluorescence inclusion, the ground truth (GT) of $\mu_a$ and $\mu_s'$ values are $\text{0.1 mm}^{\text{-1}}$ and $\text{1.0 mm}^{\text{-1}}$ respectively. A mismatched initial guess of $\mu_s'$ was assumed for FMT reconstruction. The $\mu$-adaption module successfully corrects $\mu_s'$ with an average error of 0.55\%, leading to significantly improved FMT reconstruction results compared to non-adaptive NeuFMT.
\textbf{(b)} Convergence curve of $\mu_s'$ given more initial $\mu_s'$ values ranging from 0.5$\times$ to 2$\times$ GT value under the same setting of (a).
\textbf{(c)} Convergence curve of $\mu_s'$ value across all cases with different $\mu_a$ values, given initial $\mu_s' = \text{1.5 mm}^{\text{-1}}$.
\textbf{(d-f)} Effective test for the correction of the absorption coefficient ($\mu_a$).
\textbf{(d)} For the same phantom in (a), a (mismatched) $\mu_a$ initial value is assumed for FMT reconstruction. The $\mu$-adaption module successfully corrects $\mu_a$ with an average error of 1.5\%, leading to significantly improved FMT reconstruction results. 
\textbf{(e)} Convergence curve of $\mu_a$ given more initial $\mu_a$ values ranging from 0.5$\times$ to 2$\times$ GT value under the same setting of (d).
\textbf{(f)} Convergence curve of $\mu_a$ value across all cases with different $\mu_s'$ values, given initial $\mu_a = \text{0.15 mm}^{\text{-1}}$.
\textbf{(g-h)} Effective test for the correction of both absorption and diffusion coefficients.
\textbf{(g)} For the same phantom in (a), given mismatched initial $\mu_a = $ 1.5 and $\mu_s' = $ 0.15, both of which are 50\% higher than GT, the $\mu$-adaption module successfully corrects both $\mu_a$ and $\mu_s'$ at the same time.
\textbf{(h)} Convergence curve given the iteration process in (g) for both optical coefficients.
}

\end{figure*} 

\section{Results}

\subsection{Convergence Verification of Optical Property Adaptive Optimization}
\label{sec:convergence_verification}

To validate our proposed optical property adaptation mechanism, we conducted a series of numerical experiments on a digital phantom.
The primary objective was to demonstrate that the $\mu$-adaption module can robustly recover absorption $\mu_a$ and $\mu_s'$ coefficients from severely erroneous initial estimates, thereby enabling high-fidelity fluorescence reconstruction.

We constructed a homogeneous digital slab phantom of size $\text{55} \times \text{55} \times \text{15 mm}^3$ featuring an `S'-shaped fluorescent target at its geometric center.
The ground-truth optical properties were set to $\mu_a^{\text{true}} = \text{0.1 mm}^{-1}$ and $\mu_s'^{\text{ true}} = \text{1.0 mm}^{-1}$.
To rigorously test the correction capability, we initialized the reconstruction with significantly mismatched scattering coefficients ($\mu_s'^{\text{ init}} = \text{0.5, 1.5, 2.0 mm}^{-1}$) while keeping the absorption coefficient fixed at its true value for this specific test.

As summarized in Fig.\,2 (a), without the $\mu$-adaption module, the reconstructed fluorescence distributions suffered from substantial errors in both depth localization and morphological accuracy.
The mismatched $\mu_s'$ value led to an incorrect model of photon propagation, which directly translated to distorted reconstructions.
In contrast, when the $\mu$-adaption module was enabled, the proposed $\mu$NeuFMT successfully recovered fluorescence distributions that closely matched the ground truth obtained with the correct optical properties.
Crucially, the evolution of the scattering coefficient during optimization, plotted in Fig.\,2 (b), demonstrates the module's robust convergence.
Regardless of the initial value—whether halved or doubled—the estimated $\mu_s'$ reliably converged to the neighborhood of the true value ($\text{1.0 mm}^{-1}$).

A similar convergence behavior was observed for the absorption coefficient $\mu_a$.
When initialized with biased values ($\mu_a^{\text{init}} = \text{0.05, 0.15, 0.2 mm}^{-1}$) against a fixed true $\mu_s'$, the $\mu$-adaption module corrected $\mu_a$ with an average error of only 1.5\% (Fig.\,2\,(d)), and its convergence curves are shown in Fig.\,2\,(e).
This convergence robustness was maintained across a range of true $\mu_a$ and $\mu_s'$ values, confirming the general efficacy of our alternating optimization strategy in mitigating parameter coupling (Fig.\,2 (c),\,(f)).

We further extended the validation to the most challenging scenario, where both optical parameters were simultaneously initialized with substantial errors.
In this experiment, the initial values for both $\mu_a$ and $\mu_s'$ were set to 50\% higher than their GT values.
Remarkably, even under these highly adverse conditions, the $\mu$-adaption module successfully drove both parameters towards their true values.
The convergence trajectories remained stable and ultimately reached a neighborhood of the true parameter pair. These results collectively verify that the integrated $\mu$-adaption module empowers $\mu$NeuFMT to jointly recover accurate optical properties and fluorescence distributions even from profoundly inaccurate initial guesses, effectively addressing a critical source of ill-posedness in practical FMT scenarios.

\subsection{Numerical Simulations}
\label{sec:numerical_simulations}

\begin{figure} [htbp]
\label{fig:3}
   \begin{tabular}{c}
   \includegraphics[width=\columnwidth]{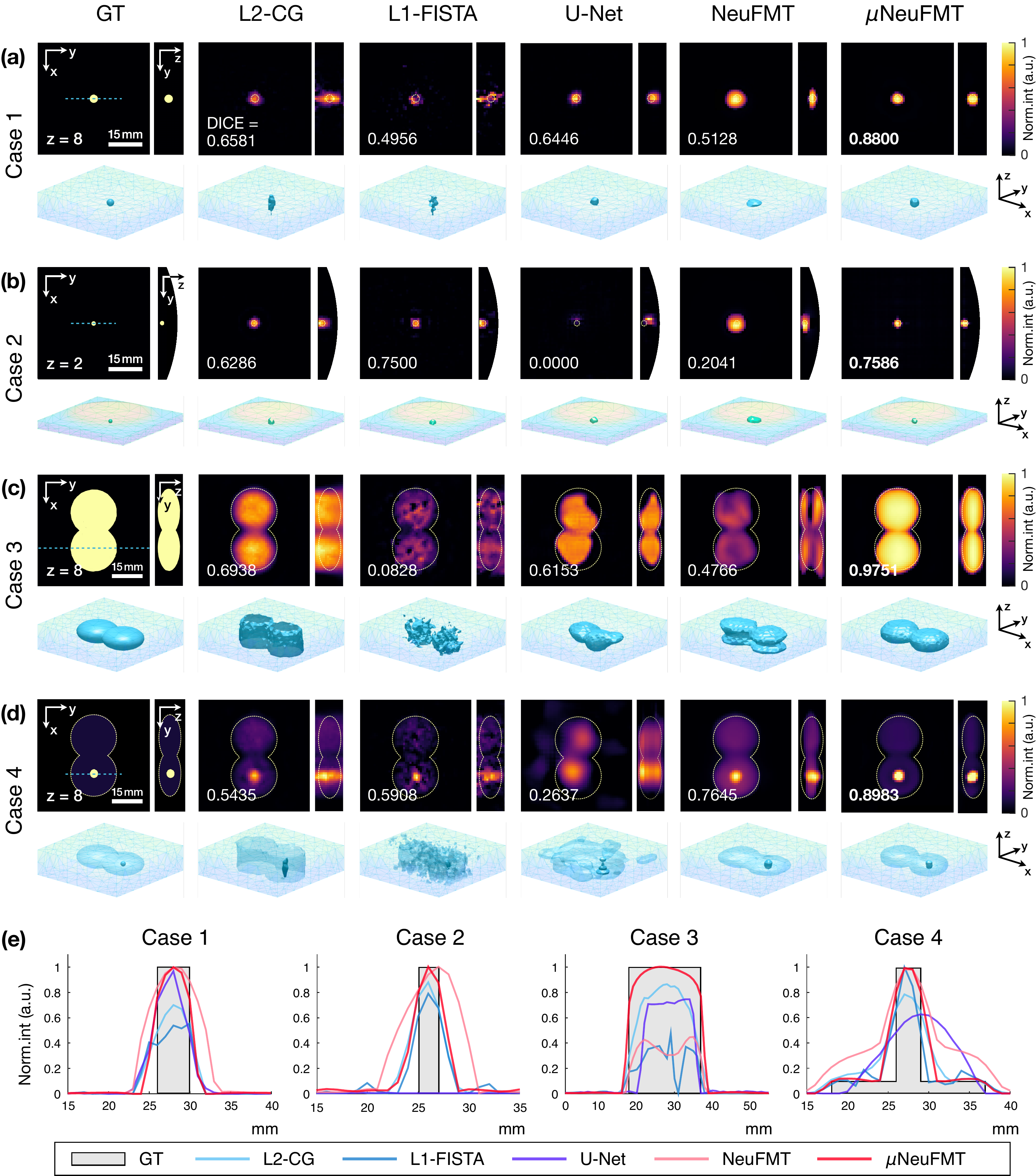}
   \end{tabular}
   \caption{ 
Results of numerical phantom simulations.
\textbf{(a)–(d)} Ground truth (GT) and FMT reconstruction results from five methods ($L_2$-CG, $L_1$-FISTA, U-Net, direct NeuFMT, and $\mu$NeuFMT) across four numerical cases. For each method, two cross-sections (along XY- and XZ-planes), a 3D isosurface, and the Dice coefficient are displayed.
\textbf{(e)} Intensity profiles along the dashed lines in the XY slices are compared for the four cases.}
\end{figure} 

To comprehensively evaluate the performance of the proposed $\mu$NeuFMT under controlled conditions, we designed four distinct numerical phantoms (Fig.\,3\,(a)\,--\,(d) respectively) to systematically increase the reconstruction challenge:
\begin{itemize}
    \setlength{\itemsep}{0pt}
    \item {Case 1 - single small target}: a 3 mm-diameter spherical fluorescence target embedded in a homogeneous cuboid phantom ($\text{55} \times \text{55} \times \text{15 mm}^3$), serving as a baseline for evaluating localization and quantitative accuracy of a simple, isolated inclusion.
    \item {Case 2 - complex surface geometry}: a 2 mm-diameter spherical target inside a phantom with a curved surface geometry, featuring a spherical cap top (base: $\text{50} \times \text{50} \times \text{4}$ mm, cap height = 5 mm, total height = 9 mm). This case tests the method's robustness to non-planar surface topography.
    \item {Case 3 - complex target shape}: a large peanut-shaped fluorescence region with uniform concentration embedded in a standard cuboid phantom same as Case 1. This scenario assesses the capability to recover intricate morphological details beyond simple spheres.
    \item {Case 4 - multi-target with background}: the same peanut-shaped region as in Case 3, but with the addition of a small, high-concentration spherical target inside. This represents a highly realistic and challenging condition, testing the ability to resolve targets of different concentrations and shapes against a noisy background.
\end{itemize}
The ground-truth optical properties for all simulated measurements in this test were set to $\mu_a =$ 0.1 mm$^{-1}$ and $\mu_s' =$ 1.0 mm$^{-1}$, and a 5\% gaussian noise is added to all simulated measurements. For each case, we compared our $\mu$NeuFMT against four established methods: conventional Tikhonov-regularized conjugate gradients ($L_2$-CG) \cite{kawata2007constrained} and $L_1$-regularized non-negative least squares ($L_1$-FISTA) \cite{beck2009fast}, a pretrained supervised U-Net, and our baseline method NeuFMT without $\mu$-adaption. The U-Net was trained on a large, synthetically generated dataset of 5,488 samples. For each sample, the input was a boundary measurement simulated using the DE-based forward model under the ground-truth properties in a cuboid phantom ($\text{55} \times \text{55} \times \text{15 mm}^3$), and the label was the corresponding 3D fluorescence map of 1\,--\,2 ellipsoids with random sizes and spatial positions. To clearly demonstrate the effect of the $\mu$-adaption module, we focused on adapting $\mu_s'$ while fixing $\mu_a$ at its true value during reconstruction. For all methods, the initial $\mu_s'$ was set to an erroneous value of 1.2 mm$^{-1}$, which is 20\% higher than the ground truth.

The results presented in Fig.\,3\,(a)\,--\,(d) demonstrate the consistent superiority of $\mu$NeuFMT across all four numerical cases.
We observe that the conventional methods were adversely affected by the optical property mismatch: $L_2$-CG reconstructions were characteristically blurred and exhibited significant depth errors, such as full-depth penetration in Cases~3 and 4 and depth-wise expansion in Case 1.
$L_1$-FISTA frequently introduced non-physical, fragmented artifacts, such as staircase patterns in Case 1 and hollow structures in Case~3, while it also suffered from severe depth inaccuracy.
The supervised U-Net, though performed well in Case 1, showed limited generalization beyond its training data, as seen in its localization failure in the complex surface geometry of Case~2 and noticeable shape distortion in the unseen peanut-shaped target of Case 3 and 4.
The direct NeuFMT provided better depth prediction than the traditional methods but remained sensitive to the erroneous $\mu_s'$, resulting in substantial shape expansion in the $x$-$y$ plane (Cases 1, 2) and axial splitting (Case 3).
In contrast, $\mu$NeuFMT successfully rectified the optical properties and overcame these limitations.
It achieved the most accurate spatial localization and shape fidelity, as evidenced by its 3D isosurfaces tightly conforming to the ground truth in both simple and complex targets, and it clearly resolved both the small target and the complex background in Case 4.
Quantitatively, $\mu$NeuFMT consistently achieved the highest Dice coefficients, and its intensity profiles (Fig.\,3\,(e)) most closely matched the ground truth in terms of peak location, width, and background suppression.

\subsection{Phantom Experiments}
\label{sec:phantom_experiments}

We further evaluated the performance of all methods using physical phantom experiments to assess their practicality under real-world conditions, including measurement noise.
The phantoms were fabricated from silicone, with their optical properties calibrated to be approximately $\mu_a = \text{0.0055 mm}^{-1}$ and $\mu_s' = \text{1.1 mm}^{-1}$ at 680 nm.
Fluorescent targets were embedded within the phantoms to create the four cases previously defined in the simulations.
The initial value of $\mu_s'$ was set identically for all reconstruction methods as the calibrated value, and also the U-Net was retrained accordingly.
Consistent with the simulation study, we focused on the adaptive recovery of the reduced scattering coefficient $\mu_s'$.

The reconstruction results are summarized in Fig.\,4.
Under these challenging real-world conditions, the limitations of the baseline methods became evident: conventional approaches ($L_2$-CG, $L_1$-FISTA) produced blurry or fragmented reconstructions with significant depth errors, while the supervised U-Net, susceptible to domain shift and noise, failed to generalize, resulting in complete localization failures (Case 1) or severely distorted shapes (Cases 3 and 4).
In contrast, the INR-based frameworks, NeuFMT and $\mu$NeuFMT, demonstrated markedly superior robustness, consistently yielding physically plausible and continuous reconstructions where other methods failed.
A detailed comparison between the two reveals that $\mu$NeuFMT provided critical refinements toward enhanced physical consistency.
In Cases 1 and 3, $\mu$NeuFMT achieved better depth sensitivity and more uniform fluorophore distribution, more faithfully representing the homogeneous nature of the true targets.
In the complex geometry of Case~2, it uniquely achieved accurate localization, whereas direct NeuFMT produced a solution with an aberrant depth shift.
For the multi-target scenario in Case 4, $\mu$NeuFMT reconstructed the background region with improved morphological accuracy.
Although the performance gap in quantitative metrics such as the Dice coefficient was occasionally narrow, which reflects the inherent robustness of the shared INR architecture, $\mu$NeuFMT provided consistent, subtle refinements, particularly in depth accuracy. The phantom experiments confirm that the adaptive physical correction mechanism in $\mu$NeuFMT can enhance reconstruction fidelity under practical conditions where the exact coefficient in most real experiments is uncertain.

\begin{figure} [ht]
\label{fig:4}
   \begin{tabular}{c}
   \includegraphics[width=\columnwidth]{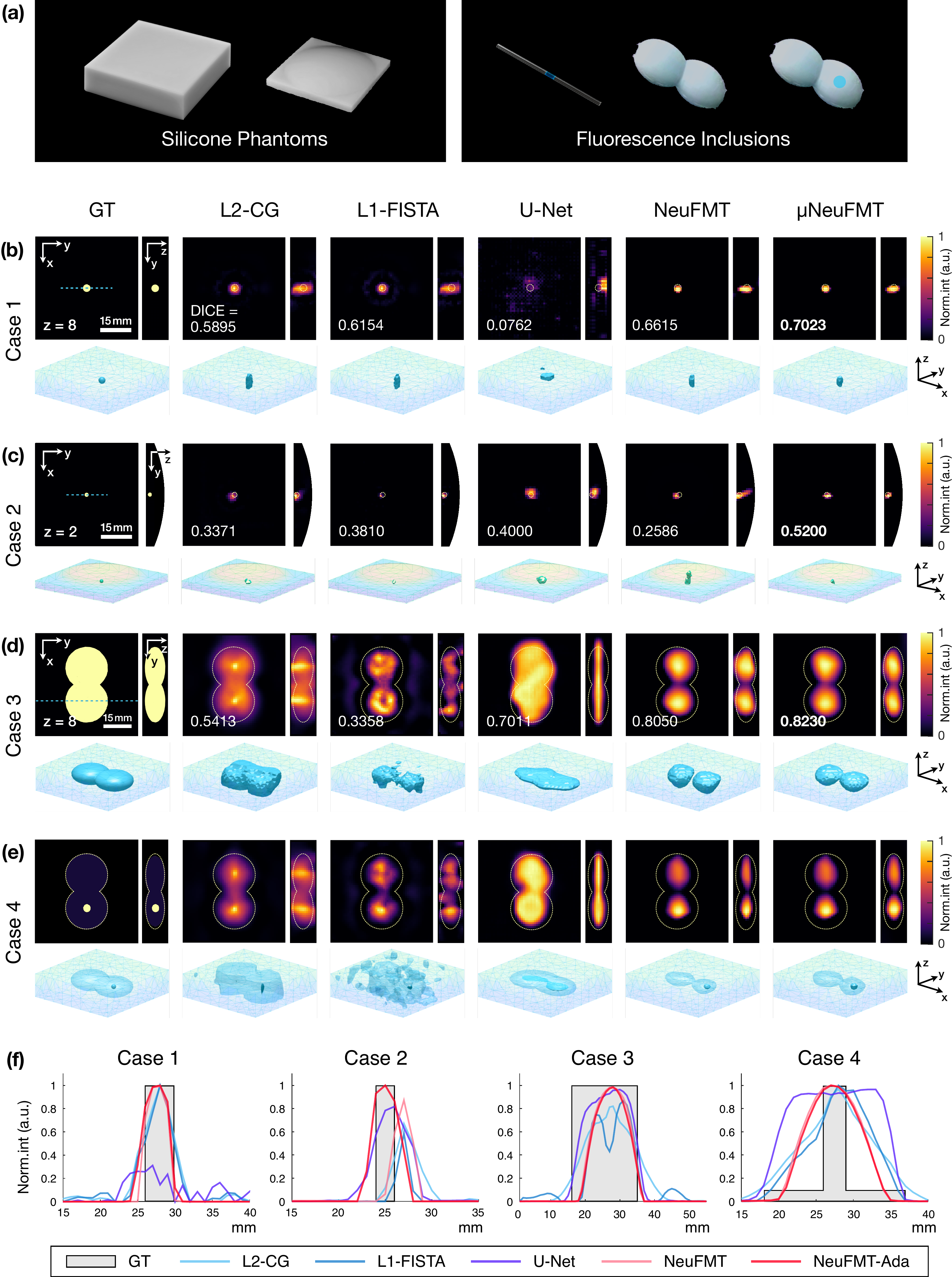}
   \end{tabular}
   \caption{ 
Results of real phantom experiments.
\textbf{(a)} Photograph of the real silicone phantoms and fluorescence inclusions.
\textbf{(b)–(e)} Ground truth (GT) and FMT reconstruction results from five methods ($L_2$-CG, $L_1$-FISTA, U-Net, direct NeuFMT, and $\mu$NeuFMT) across four numerical cases. For each method, two cross-sections (along XY- and XZ-planes), a 3D isosurface, and the Dice coefficient are displayed.
\textbf{(f)} Intensity profiles along the dashed lines in the XY slices are compared for the four cases.}
\end{figure} 

\subsection{In Vivo Lymph Node FMT Imaging Experiments}

We performed an \emph{in vivo} imaging experiment on mice.
Following the protocol described in Subsection \ref{sec:invivo_setup}, Cy5 fluorescent dye accumulated primarily at the sentinel lymph nodes (LN), with some residual signal at the vein (Fig.\,5\,(a)).
Our task was to reconstruct the spatial location and shape of the sentinel lymph nodes.

The 3D FMT reconstruction results from all methods are presented in Fig.\,5\,(b).
The conventional $L_2$-CG method produced a reconstruction where the volume of LN-3 was overestimated, and the signals from LN-1, LN-2, and the vein merged into a single, indistinguishable cluster.
The $L_1$-FISTA reconstruction was highly fragmented, making it impossible to discern the shape or count of the lymph nodes accurately.
Also, the baseline NeuFMT failed to produce a physiologically plausible reconstruction.
In contrast, $\mu$NeuFMT was the only method that successfully and distinctly resolved all three lymph nodes (LN-1, LN-2, LN-3) and the vein, clearly identifying each as separate, localized regions of fluorescence.

For quantitative validation, the mouse was sacrificed, and the lymph nodes were excised to be imaged by \emph{ex vivo} FRI (Fig.\,5\,(c)).
The reconstructed target lengths from $\mu$NeuFMT showed the closest agreement with the physical measurements taken from the \emph{ex vivo} lymph nodes, as visualized by the intensity profiles in Fig.\,5\,(c).
These results demonstrate that $\mu$NeuFMT provides a definitive advantage in practical \emph{in vivo} scenarios, enabling robust and accurate target segmentation and localization under complex clinically related scenarios, such as fluorescence guided surgery.

\begin{figure} [htbp]
\label{fig:5} 
   \begin{center}
   \begin{tabular}{c}
   \includegraphics[width=\columnwidth]{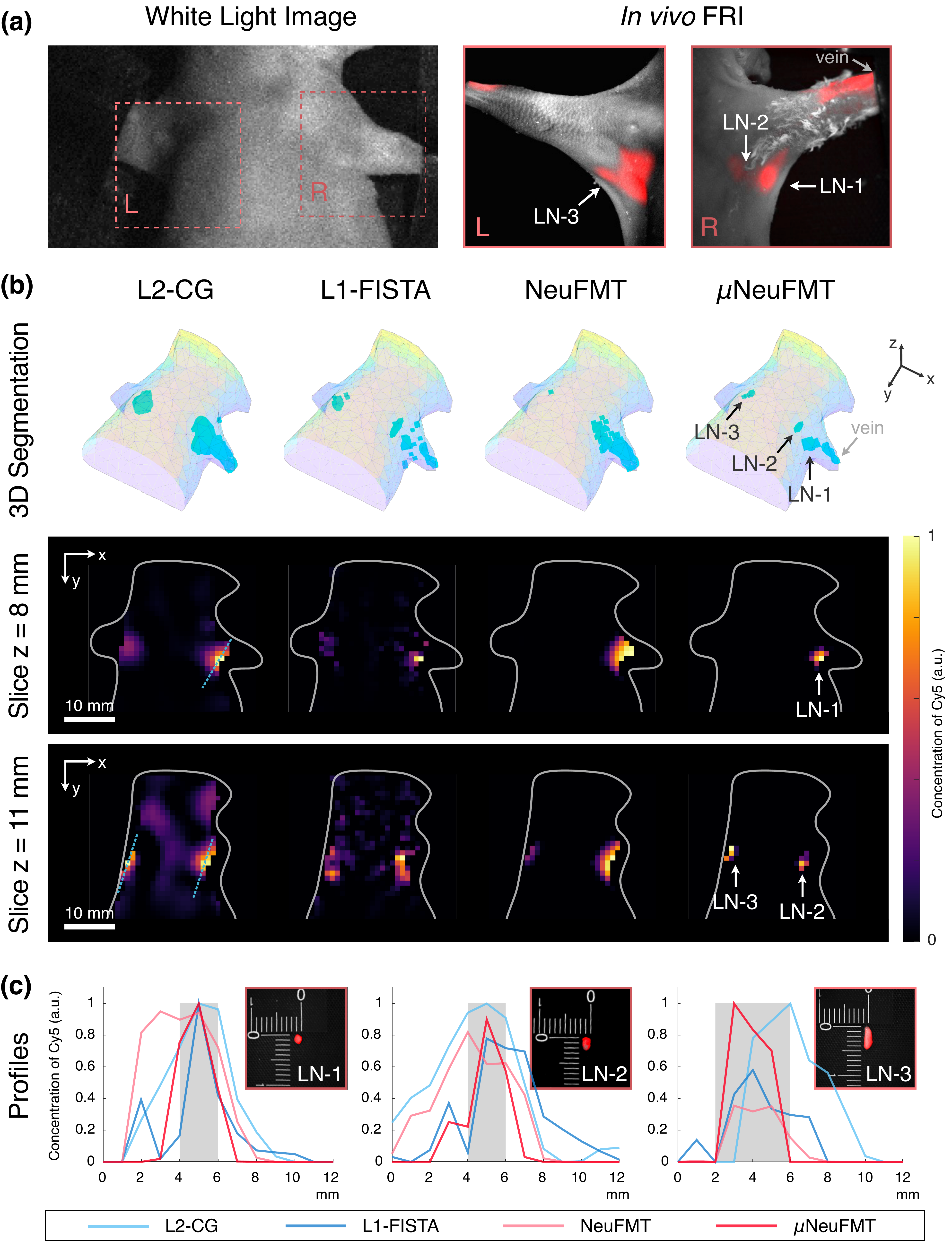}
   \end{tabular}
   \end{center}
   \caption{ 
\emph{In vivo} lymph node FMT imaging and reconstruction results.
\textbf{(a)} The white-light image and preview via 2D fluorescence reflectance imaging (FRI) overlaid on the white-light image with the red aera indicating the concentrated distribution of Cy5 probe on the cites of three lymph nodes (LN-1, LN-2, LN-3) and the vein.
\textbf{(b)} Comparison of FMT reconstruction results from different methods. The 3D segmentation, along with two slices at $z$ = 8 and 11 mm, are displayed.
\textbf{(c)} Intensity profiles along the dashed lines in b. are compared for each \emph{ex vivo} images of lymph nodes after sacrificing the animal. The gray zone represents the estimated size of LN, based on the \emph{ex vivo} fluorescence images.
}
\end{figure} 

\section{Discussion and Conclusion}

In this study, we introduced $\mu$NeuFMT, an optical-property-adaptive FMT reconstruction framework that integrates INR with an explicitly differentiable FEM-based light propagation model. By unifying a continuous fluorescence field parameterized by an INR and a physics-based forward model amenable to end-to-end optimization, $\mu$NeuFMT addresses two fundamental limitations of existing FMT reconstruction techniques: their inability to efficiently adapt to complex and heterogeneous fluorescence distribution, and their reliance on fixed, often inaccurate, optical properties, which significantly exacerbates the inherent ill-posedness of the inverse problem.

Unlike supervised deep learning approaches which are constrained by the diversity and scale of training datasets and often fail under domain shift, $\mu$NeuFMT operates in a self-supervised, physics-informed regime. It does not require paired training data; instead, it directly minimizes the discrepancy between measured and simulated boundary fluxes. This not only eliminates the need for extensive experimental datasets but also ensures generalizability across different imaging geometries and target configurations. The continuous, coordinate-based representation provided by the INR further ensures that our reconstructions are free from discretization artifacts and inherently resolution-agnostic.

A central contribution of our work is the introduction of an optical property adaptation mechanism, which enables the dynamic optimization of absorption and scattering coefficients ($\mu_a$, $\mu_s'$) concurrently with the fluorescence distribution. Our framework demonstrates that the integration of a differentiable physics-based forward model is the key to enabling this self-supervised adaptation. The FEM-based light propagation model is not merely a component of the loss function, but an active and adaptive part of the inverse solver. This stands in contrast to both traditional iterative methods, which hold optical properties static and thus aggravate the problem's ill-posedness, and conventional end-to-end deep learning approaches, which struggle to generalize beyond their training data. The adaptivity of our model is thus its core strength, providing a principled way to refine the physical model itself based on the measured data, which is crucial for practical applications where precise prior knowledge of tissue optics is unavailable.

Our results from numerical simulations, phantom experiments, and in vivo mouse imaging collectively demonstrate that $\mu$NeuFMT achieves superior reconstruction fidelity compared to both conventional ($L_2$, $L_1$) and deep learning-based (U-Net) benchmarks. In particular, the method consistently localized fluorescent targets with higher spatial accuracy, improved contrast, and more realistic morphological features—especially under challenging conditions involving background fluorescence and complex geometry.

Despite these promising results, several aspects of $\mu$NeuFMT merit further investigation. First, the current implementation assumes a homogeneous optical background—a common simplification in FMT studies. Extending the model to incorporate spatially varying optical properties represents a logical next step, potentially by representing $\mu_a(\br)$ and $\mu_s'(\br)$ with additional INRs, though this would significantly increase the complexity of the inverse problem. 
Nevertheless, the proposed $\mu$NeuFMT can be validated or even combined with time-domain near-infrared spectroscopy (fNIRS) \cite{yamada2019time, ban2022kernel} or diffuse optical tomography (DOT) \cite{pifferi2016new} techniques, where the complex optical properties related with absorption and scattering can disentangled and calibrated.
More importantly, the integration of $\mu$NeuFMT with emerging imaging technologies holds particular promise. The inherent adaptivity of our framework is well-suited to the distinct scattering and absorption landscapes of NIR-II fluorescence tomography \cite{wang2024vivo, zhang2024nir}, where our optical property optimization could be crucial for accurately modeling the increased photon penetration and achieving higher-resolution reconstruction at depth. Furthermore, the framework can be naturally extended to hyperspectral FMT, using a single INR to output both fluorophore concentration and spectral signature, thereby disentangling multiple probes or accounting for wavelength-dependent optical properties. Finally, translating $\mu$NeuFMT into clinical settings, such as intraoperative fluorescence-guided surgery \cite{mieog2022fundamentals, wang2023fluorescence}, represents an exciting frontier where its adaptability could overcome patient-specific variations in tissue optics.

In conclusion, by harmonizing a differentiable physics model with a continuous INR, $\mu$NeuFMT provides a robust, self-supervised, and adaptive framework for FMT reconstruction. It effectively mitigates the ill-posedness inherent to inverse problems with uncertain parameters, setting a new paradigm for practical, high-resolution fluorescence molecular imaging. We believe this `physics-in-the-loop' INR approach offers a generalizable template for tackling a wide range of tomographic and inverse problems in biomedical imaging.

\bibliographystyle{IEEEtran}
\bibliography{references}

\end{document}